\documentstyle[epsfig,12pt]{article}

\newcommand{\lbl}[1]{\label{eq:#1}}
\newcommand{\rf}[1]{(\ref{eq:#1})}
\newcommand{\be}{\begin{equation}}
\newcommand{\en}{\\[2mm]\end{equation}}
\newcommand{\bea}{\begin{eqnarray}}
\newcommand{\ena}{\end{eqnarray}}
\newcommand{\NP}[1]{Nucl.\ Phys.\ {\bf #1}}
\newcommand{\ZP}[1]{Z.\ Phys.\ {\bf #1}}

\newcommand{\PL}[1]{Phys.\ Lett.\ {\bf #1}}
\newcommand{\NC}[1]{Nuovo Cimento {\bf #1}}
\newcommand{\AN}[1]{Ann. Phys. {\bf #1}}

\newcommand{\PRev}[1]{Phys.\ Rev.\ {\bf #1}}
\newcommand{\PRL}[1]{Phys.\ Rev.\ Lett.\ {\bf #1}}

\newcommand{\lapprox}{%
\mathrel{%
\setbox0=\hbox{$<$}
\raise0.6ex\copy0\kern-\wd0
\lower0.65ex\hbox{$\sim$}
}}
\newcommand{\gapprox}{%
\mathrel{%
\setbox0=\hbox{$>$}
\raise0.6ex\copy0\kern-\wd0
\lower0.65ex\hbox{$\sim$}
}}

\def\fpi{F_\pi}
\def\M{M_\pi}
\def\mpi{M_\pi}
\def\F{{\bf F}}
\def\W{{\bf W}}
\def\WW{{\hat W}}
\def\Z{{\bf Z}}
\def\a{{\bf a}}
\def\b{{\bf b}}
\def\h{{\bf h}}
\def\chpt{$\chi$PT}
\def\gchpt{$G\chi$PT}
\def\pp{$\pi\pi$\ }
\def\Im{{\rm Im}\,}
\def\intlow{\int_{4\M^2}^{E^2}\, }
\def\inthi{\int_{E^2}^\infty\,   }
\def\GeV{{\rm GeV}}
\def\today{\ifcase \month\or
  January\or February\or March\or April\or May\or June\or
  July\or August\or September\or October\or November\or December\fi
  \space\number\day,\space
  \number\year }
\begin{document}
\rightline{IPNO/TH 95-64}
\rightline{PURD-TH-95-10}
\vskip 2 true cm
\centerline{\Large\bf Determination of Two-Loop $\pi\pi$ }
\centerline{\Large\bf Scattering Amplitude Parameters}
\vskip 2.5 true cm
\begin{center}
{\bf M. Knecht, B. Moussallam, J. Stern}\\
{\sl Division de Physique Th\'eorique
\footnote{Unit\'e de Recherche des Universit\'es Paris XI et Paris VI
associ\'ee au CNRS.}, Institut de Physique Nucl\'eaire\\
F-91406 Orsay Cedex, France} \\

\medskip
{\bf N.H. Fuchs}\\
{\sl Department of Physics, Purdue University\\
West Lafayette, IN 47907, USA}\\
\end{center}
\vskip 2 true cm
\centerline{\large\bf Abstract}
 
The chiral expansion of the \pp amplitude to the order of two loops
was expressed in terms
of six independent parameters in a previous
paper: four of these are shown here to satisfy sum rules. Their
derivation, where crossing symmetry plays a key role,
is explained. Their convergence properties are studied and
their practical evaluation, in terms of the available data on \pp
phase shifts above 0.5 GeV, is discussed. Below 0.5 GeV, the
chiral amplitude itself is employed, such that the parameters 
are determined in a self-consistent way. Some care is devoted
to the estimate of the errors.
 
\vfill\eject
\leftline{\large\bf 1. Introduction}
\bigskip

In a previous paper \cite{kmsf} ( referred to as I in the sequel),
we showed that the \pp elastic scattering amplitude to the order of
two loops in the chiral expansion takes the form of an analytic 
expression that depends on six parameters ($\alpha,\ \beta,\ 
\lambda_1,\ \lambda_2,\ \lambda_3,\ \lambda_4$) which are not determined
by chiral symmetry. In the present work, the possibility of determining
these parameters from experiment will be discussed. It will be shown
that four of them, $\lambda_1,...,\lambda_4$, can be inferred with 
rather good accuracy from existing \pp scattering data at energies 
$\sqrt{s}\ge 500$ MeV. This determination is merely based 
on very general properties of the scattering amplitude and it is
completely model independent. Before entering into the details
of the subject \footnote{
A brief account of the results of 
the present paper has already been presented in I
(sec. 4).}
a few comments on the low-energy constants characterizing  higher 
orders in the low-energy expansion may be useful.

Upon extending the program of the chiral perturbation theory (\chpt) 
beyond one-loop order \cite{gl84,gl85}, one faces the problem of
the proliferation of low-energy constants: at order $O(p^6)$ 
there are more than a hundred new unknown parameters \cite{fearing}. 
At one-loop level, the predictive power of \chpt\ resides in the fact
that the same (combinations of) low-energy constants often appear in 
different observables. This fact, of course, reflects chiral Ward 
identities and unitarity and it is encoded in the effective lagrangian.
However, at two-loop order, eliminating 
unknown parameters in the same way by 
comparing different observables becomes problematic in practice, 
though not impossible in principle. In spite of this difficulty,
successful two-loop calculations do exist \cite{bgs}, in which the
influence of the new $O(p^6)$ constants remains rather limited, at least
in a particular kinematical region.

In general, it seems plausible
that at $O(p^6)$ accuracy, each 
experimentally relevant process will be described by its own, specific
set of low-energy constants, which have to be determined from 
considerations operating beyond the strict framework of the low-energy
expansion. Models aiming at such a determination have been proposed,
and already applied in the past both at the $O(p^4)$ \cite{egpr,eglpr}
and $O(p^6)$ \cite{bgs} levels. They incorporate low-lying meson 
resonances into the effective lagrangian, thereby extending its domain
of applicability. The unknown low-energy constants
are then obtained by fitting 
the resonance parameters and couplings first, and then integrating out the
resonance degrees of freedom (expanding in inverse
powers of their masses). While this procedure is essentially unique
and presumably rather accurate in the case of vector or axial-vector
resonances \cite{eglpr}, it is less convincing as far as scalar 
exchanges are concerned: i) first, incorporating spin 0 resonances 
into ${\cal L}_{eff}$ in a way consistent with chiral symmetry
is not a unique procedure. For obvious kinematical reasons, the criterion
of matching with QCD at short distances \cite{eglpr} is less stringent
than in the case of spin-1 exchanges. Too many terms survive this criterion,
which even requires the presence of scalar couplings with a non-minimal
number of derivatives \cite{ms} (which were not considered in \cite{egpr}).
ii) Next, it has been suggested that the spectrum and couplings of the scalar 
mesons are subject to particularly strong distortions due to unitarity
and final-state interactions \cite{tornqvist}. It may be misleading to
describe the $0^{++}$ channels by a set of quasi-real poles or,
equivalently, to identify the resonance masses and couplings in the
extended effective lagrangian with the corresponding entries in the PDG
tables \cite{pdg}. For all these reasons, a more realistic and less model
dependent method of determination of low-energy constants is needed, 
especially in the symmetry breaking sector which is particularly
sensitive to scalar exchanges.

In this paper, such an unambiguous method is worked out in the 
case of elastic \pp scattering. It may be viewed as a generalization
of the method of chiral sum rules for various two-point 
functions \cite{chisum,ggpp}, which has been already used to determine
some $O(p^6)$ constants in connection with the reaction $\gamma\gamma\to
\pi^0\pi^0$ \cite{ggpp}. 
The method makes use of analyticity, unitarity and crossing symmetry
to relate the existing experimental information on the \pp
scattering amplitude at medium energies, $0.5\ {\rm GeV}\le\sqrt{s}
\le1.9\ {\rm GeV}$, to its behaviour in the threshold region, 
where \chpt\ applies. 
The comparison yields six sum rules for the six unknown
parameters $\alpha,\ \beta,\ \lambda_1,...,\lambda_4$ which define the
perturbative \pp amplitude up to two loops. Four of these sum rules 
converge rapidly enough to be practically insensitive to the poorly
known high-energy part of the \pp amplitude. In the absence of 
information on this part, the remaining two sum rules are of little use
in practice. For this reason, our method merely serves to fix the
constants $\lambda_1,...,\lambda_4$, leaving the parameters $\alpha$
and $\beta$ undetermined. This, of course, is reminiscent of the
derivation of the Roy equations \cite{roy,leguillou} for the \pp
scattering amplitude: here, as well, after imposing crossing symmetry
and fixed-t analyticity, one remains with two arbitrary subtraction 
constants \cite{roy,leguillou}. The parallel with the Roy equations will
be further developed in sec. 2, where the derivation of the sum rules is 
presented in details.

The \pp amplitude is a rare example of a low-energy observable
which is sensitive to the mechanism of spontaneous chiral symmetry
breaking, in particular to the strength of quark antiquark condensation
in the QCD vacuum \cite{fss,ssf}.
Indeed, elastic \pp scattering 
provides the best way of testing 
the standard postulate, according to which
the quark condensate
\be
B_0=-\lim_{m_q\to0} {1\over F_\pi^2} <\bar q q>
\en
is large enough (c.f. $B_0\sim 1.5$ GeV ) 
to dominate the response to the perturbation by light quark masses
\cite{gor}. 
This hypothesis is at the basis of the standard
\chpt\ \cite{gl84,gl85}, which explicitly {\it assumes} that the 
GOR relation $2\hat m B_0=M_\pi^2$ is not violated by more than a few
percent \cite{gl84}. This assumption has not yet been tested experimentally, 
and it is not a necessary prerequisite for a
consistent chiral expansion: indeed, it was shown in \cite{fss,ssf,ggpp} 
that a systematic,  
more general expansion (\gchpt) can be developed, which neither
requires $B_0$ to be large, nor the expansion of $\M^2$ to be dominated
by $2\hat m B_0$ (see \cite{daphne} for a recent review). The two-loop
expression of the \pp amplitude displayed in I holds in full
generality, independently of any particular mechanism of chiral
symmetry breaking. The strength of quark condensation merely affects
the values of the parameters $\alpha,\ \beta,\ \lambda_1,...,
\lambda_4$, but not the analytic form of the amplitude. The standard and
the generalized \chpt\ differ in the expansion of these parameters
in terms of quark masses, chiral logarithms and the renormalized (quark mass
independent) constants of ${\cal L}_{eff}$. This expansion was 
worked out very recently \cite{bij} up to chiral order six in the
standard \chpt.
This kind of information
is particularly useful for the two parameters $\alpha$ and $\beta$, 
the expansion of which start at the lowest $O(p^2)$ order
of the \chpt, and higher order contributions  are expected to remain
relatively small. In the standard \chpt, both $\alpha$ and $\beta$
are predicted to be close to 1 \cite{gl84} (within $10-20$\%, say), while
in the alternative situation of a small condensate the value of $\alpha$
would be $2-3$ times larger, depending on the actual value 
of the quark mass ratio $m_s/\hat m$, which remains a free 
parameter in the \gchpt. In all
cases the deviation of $\beta$ from 1 should remain small: for more
details, see I. In contrast, the expansions of the parameters 
$\lambda_1,\ \lambda_2$ and $\lambda_3,\ \lambda_4$ start at order
$O(p^4)$ and $O(p^6)$, respectively. Consequently, these parameters are
more sensitive to the unknown $O(p^6)$ constants of ${\cal L}_{eff}$ and
their values are thus expected to be more difficult to estimate on the basis
of a \chpt\  calculation alone. It is gratifying that the experimental 
\pp data at medium energies allow to determine 
$\lambda_1,...,\lambda_4$, and it is significant that the same data
leave $\alpha$ and $\beta$ undetermined. Our analysis further
suggests that the  values of $\lambda_1,...,\lambda_4$ are practically
insensitive to the strength of quark condensation. We checked that by 
allowing $\alpha$ to vary in a rather wide range (from 1 to 4), 
the calculated values of the $\lambda_i$'s were only affected at 
the level of a few percent.
In particular, the
values we obtain should be used within the framework of the 
standard {\chpt}  \cite{bij} (see sec. 3.5).

In order to disentangle the large and small condensate alternatives, 
it would be ideal to be able to
exploit the variation of the parameters 
$\alpha,\ \beta,\ \lambda_1,...,\lambda_4$ as functions of quark masses
and chiral logarithms as predicted in the standard and generalized
{\chpt}, respectively. Unfortunately, this kind of theoretical information
is not suitable for a direct experimental test, since Nature has already
made its choice of quark masses. On the other hand, the forthcoming
high-precision data near threshold \cite{franzini,dirac} should allow
to measure the parameter $\alpha$, for which the alternatives in question
differ by a factor $2-3$. Obviously, determining all six parameters
of the two-loop amplitude from a fit to the low-energy data appears
practically hopeless. For this reason, the model independent
determination of the parameters $\lambda_1,...,\lambda_4$ reported
in this paper becomes of a fundamental importance. Special care is
devoted to the discussion of the experimental inputs and of the 
uncertainties in the results (sec. 3). A reliable estimate of the 
errors is indeed crucial to correctly assess the size of the expected
theoretical uncertainty in future experimental determinations of the
critical parameter $\alpha$.

\bigskip 
\leftline{\large\bf 2. Derivation of the sum rules}

\bigskip
We start by giving a brief description of the \pp amplitude to
two loops, in order to set up the definition of the parameters for
which we intend to establish the sum rules. We will neither repeat the
derivation nor even give the complete formulas,
all these details can be found in I.
The $T$-matrix element for the process $\pi^a\pi^b\to\pi^c\pi^d$ is expressed
in a standard way in terms of a single function $A(s\vert t,u)$, 
symmetric in $t,u$, and where the
Mandelstam variables $s,t,u$ satisfy $s+t+u=4\mpi^2$, as
\be
{ T}^{ab,cd}(s,t,u)=A(s\vert t,u)\delta^{ab}\delta^{cd} +
A(t\vert s,u)\delta^{ac}\delta^{bd} +
A(u\vert t,s)\delta^{ad}\delta^{bc} .
\en
At two-loop order, the function $A(s\vert t,u)$ may be written as a
sum of two terms
\be\lbl{A1}
A^{2-loop}(s\vert t,u)=A^{pol}(s\vert t,u)+A^{cut}(s\vert t,u)\ ,
\en
where the former is a polynomial in the Mandelstam variables and the
latter collects the cuts of the amplitude. The polynomial is
of the third order and satisfies crossing symmetry. It may be
parametrized as
\bea\lbl{Apol}
A^{pol}(s\vert t,u)=
&&{\beta\over\fpi^2}\,\left( s-{4\mpi^2\over3} \right)
+ {\alpha\over\fpi^2}\,{\mpi^2\over3}\nonumber\\
&&+{\lambda_1\over\fpi^4}(s-2\mpi^2)^2
+{\lambda_2\over\fpi^4}\left[(t-2\mpi^2)^2+(u-2\mpi^2)^2\right] \\
&&+{\lambda_3\over\fpi^6}(s-2\mpi^2)^3
+{\lambda_4\over\fpi^6}\left[(t-2\mpi^2)^3+(u-2\mpi^2)^3\right] \ ,\nonumber
\ena
which displays the six parameters $\alpha,\ \beta,\ \lambda_1,\
\lambda_2,\ \lambda_2,\ \lambda_4$ of the two-loop amplitude.
The part containing the cuts in the complex plane
can be expressed in terms
of three functions of a single variable $\WW_a(z),\ a=0,1,2$, which are
i) analytic in the whole complex
plane in the variable $z$ except for a right-hand cut on the real axis
$[4\mpi^2,\infty]$, and ii) asymptotically bounded when $z\to\infty$,
such that $\WW_{a}/z^4\to0$ for $a=0,2$ 
and $\WW_1/z^3\to0$.
In terms of these
functions, $A^{cut}(s\vert t,u)$ can be written as
\bea\lbl{Acut}
{1\over{32\pi}}\,A^{cut}(s\vert t,u)&=&
{1\over3}\left[\WW_0(s)-\WW_2(s)\right]\nonumber\\
&+&{1\over2}\left[ 3(s-u)\WW_1(t)+\WW_2(t)\right]\nonumber\\
&+&{1\over2}\left[ 3(s-t)\WW_1(u)+\WW_2(u)\right]\ .
\ena
The fact that the amplitude to two loops has the general structure
described in the above equations \rf{A1},\rf{Apol},\rf{Acut} was
first proved in \cite{ssf}. In the subsequent paper I,
the explicit form
of the three functions $\WW_a(z)$ 
was obtained \footnote{
The functions $W_a(z)$ introduced in I differ from the functions
$\WW_a(z)$ 
given in \rf{Wa} by  polynomials which are defined up to an ambiguity described
in appendix B of \cite{ssf}. }
\be\lbl{Wa}
\WW_a(z)=\sum_{n=0}^4 w_a^n(z)\bar K_n(z)\ ,
\en
where
\be
\bar K_0(z)={1\over16\pi^2}\left\{2+\sqrt{1-4\mpi^2/z}\,\ln\left(
1-{2\over1+\sqrt{1-4\mpi^2/z}} \right)\right\}
\en
is the Chew-Mandelstam function which is already present at 
one-loop order,
and the other four functions $\bar K_1,...,\bar K_4$ are simple
combinations of $\bar K_0$ , $(\bar K_0)^2 $ and $(\bar K_0)^3$.
The set of functions $w_a^n(z)$ are third-degree polynomials
in $z$ which are tabulated in I. These polynomials
depend on four constants,
$\alpha,\ \beta,\ \lambda_1,\ \lambda_2$
which are the same (up to terms of order $O(p^8)$ in the amplitude)
as those which appear in \rf{Apol}. The dependence is cubic in
$\alpha$ and $\beta$ and linear in $\lambda_1$ and $\lambda_2$.
Finally, we note that there is a simple relation among the discontinuities
of the functions $\WW_a$ and the discontinuities  of the S and P
partial waves along the right-hand cut, $s\ge4\mpi^2$:
\be\lbl{imw}
{\rm Im} f_0^{a}(s)={\rm Im} \WW_a(s),\ a=0,2,\qquad
{\rm Im} f_1^1(s)=(s-4\mpi^2)\,{\rm Im} \WW_1(s) .
\en
This relation is actually used in the construction of the
functions $\WW_a$ in a recursive way, starting from the expression
of the partial waves at order $O(p^2)$ and using unitarity to generate
the imaginary part at a higher order. This technique was used for the
first time in \cite{lehman} to obtain the one-loop \pp amplitude
in the case $\M=0$. 

\bigskip 
\noindent{\bf 2.1 Roy dispersion relations}

\bigskip
Due to the Froissart bound, the \pp scattering amplitude 
obeys fixed $t$, twice-subtracted dispersion relations which
involve three subtraction functions.
It was shown by Roy \cite{roy} and by Basdevant {\it et al.} 
\cite{leguillou} that these three functions  are entirely determined, up
to two constants
(which, for instance, may be taken as the two scattering lengths $a_0^0$ and
$a_0^2$), once crossing symmetry is imposed. The key step in the 
derivation of the sum rules is to equate, in the low energy region, 
the perturbative expansion of the amplitude \rf{A1} \rf{Apol} \rf{Acut}
with the Roy dispersive
representation. We start by rederiving this representation in a form
which will be convenient for our purpose.
 
Consider the $s$-channel isospin $I=0,1,2$ amplitudes $F^{I}(s,t)$,
defined as 
\bea\lbl{Fdef}
F^0(s,t) &=& {1\over{32\pi}}\, \left\{ 3A(s\vert  t,u) + A(t\vert s,u) +
A(u\vert s,t)\right\} \ ,\nonumber\\
F^1(s,t) &=& {1\over{32\pi}}\, \left\{ A(t\vert s,u) -
A(u\vert s,t)\right\}\lbl{F^I} \ ,\\
F^2(s,t) &=& {1\over{32\pi}}\, \left\{ A(t\vert s,u) +
A(u\vert s,t)\right\}\ ,\nonumber
\ena
 and form a 3-vector
\be
\F(s,t)=\left(
\begin{array}{c}
F^0(s,t)\\
F^1(s,t)\\
F^2(s,t)\\
\end{array}\right).
\en
The Froissart bound allows one to write a fixed $t$ dispersion relation
with two subtractions
\be
\F(s,t)=C_{st}\big[\a_+(t)+(s-u)\a_-(t)\big]+{1\over\pi}
\int_{4\M^2}^\infty {dx\over x^2}\left( {s^2\over x-s}+
{u^2C_{su}\over x-u} \right){\rm Im}\F(x,t)\ ,
\lbl{disp}
\en
where $C_{st}$, $C_{su}$ and $C_{tu}$ are the $s-t$, $s-u$ and $t-u$
crossing matrices
\be
C_{st}=\left(
\begin{array}{rrr}
1/3 & 1   & 5/3 \\
1/3 & 1/2 &-5/6 \\
1/3 &-1/2 & 1/6 \\
\end{array}\right),\quad
C_{su}=\left(
\begin{array}{rrr}
 1/3 & -1   & 5/3 \\
-1/3 & 1/2 &  5/6 \\
 1/3 & 1/2 &  1/6 \\
\end{array}\right),\quad
C_{tu}=\left(
\begin{array}{rrr}
 1 & 0  & 0 \\
 0 & -1 & 0 \\
 0 & 0  & 1 \\
\end{array}\right) \ .
\en
We have also introduced the vectors
\be
\a_+(t)=
\left(
\begin{array}{c}
a_0(t)\\
0\\
a_2(t)\\
\end{array}\right),\qquad
\a_-(t)=\left(
\begin{array}{c}
0\\
a_1(t)\\
0\\
\end{array}
\right),
\lbl{apam}
\en
which collect three arbitrary functions of $t$. As was observed in
\cite{roy,leguillou}, the constraint of crossing symmetry reduces this
arbitrariness down to two real constants.  From \rf{disp}, $\F$
manifestly satisfies $s-u$ crossing by construction, {\it i.e.}
$\F(s,t) = C_{su}\F(u,t)$.  One must impose $s-t$ crossing symmetry,
and then $t-u$ crossing symmetry will follow automatically.
 
It proves convenient to split the
integration range into two parts, $[4\M^2,E^2]$ and $[E^2,\infty]$.
In the classic study of the Roy equations \cite{bfp} $E$ was taken to be 
rather large, $E\simeq1.5$ GeV. Here, for the purpose
of generating a form which can easily be compared with the
two-loop expression, 
we will take $E$ to be sufficiently small, so 
that in the range $4\M^2\le x\le E^2$ 
the contributions of the partial waves
with $\ell\ge2$ to the imaginary part of the amplitude 
are negligibly small as compared to the S- and P-wave contributions. 
This condition is satisfied for $E\lapprox 1$ GeV. 
For $x\le E^2$ we can write
\be\lbl{Imf}
\Im\F(x,t)=\left(
\begin{array}{c}
\Im f_0^0(x)\\
0           \\
\Im f_0^2(x)\\
\end{array}
\right) +3\Big(1+{2t\over x-4\M^2}\Big)
\left(
\begin{array}{c}
0\\
\Im f_1^1(x)\\
0\\
\end{array}
\right)\ .
\en
Inserting this into the dispersion relation \rf{disp}, there
will appear three functions
$W_a^E(z)$ which are analogous   to the
functions $\WW_a(z)$ in the two-loop expression:
\be\lbl{Wae}
W_{a}^E(z)={z^2\over\pi}\intlow{
dx\over x^2}{\Im f_0^a(x)
\over x-z}\ \ (a=0,2)\ ,\qquad
W_{1}^E(z)={z\over\pi}\intlow{
dx\over x(x-4\M^2)}{\Im f_1^1(x)
\over x-z}\ .
\en
We collect these functions into vectors
\be
\W^E_+(z)=
\left(\begin{array}{c}
W^E_0(z)\\
  0\\
W^E_2(z)\\
\end{array}\right),
\qquad
\W^E_-(z)=
\left(\begin{array}{c}
 0 \\
W^E_1(z)\\
 0 \\
\end{array}\right).
\lbl{Wpe}
\en
The amplitude $\F(s,t)$ can then be expressed as a sum of two terms
\be
\F(s,t)=\W^E(s,t)+\Z^E(s,t)\ .
\en
The first term is constructed from the integrals $W_a^E$ in order
to  satisfy crossing symmetry exactly:
\bea\lbl{WE}
\W^E(s,t)=&&\W^E_+(s)+
3(t-u)\W^E_-(s) +C_{su}[\W^E_+(u)+3(t-s)\W^E_-(u)] \nonumber\\
&&+C_{st}[\W^E_+(t)+3(s-u)\W^E_-(t)]\ .
\ena
The second term, $\Z^E(s,t)$, collects what is left
from the original dispersive representation. Remarkably,
it can be expressed solely in terms
of integrals over the high-energy range $[E^2,\infty]$ and of three
arbitrary functions of $t$. This is achieved by redefining the original
subtraction functions $\a_+(t)$ and $\a_-(t)$ into new ones
$\b_+(t)$ and $\b_-(t)$ (which are equally arbitrary for the moment),
collecting into them as many terms as possible: 
\bea\lbl{subrel}
&&\b_+(t)=\a_+(t)-\W^E_+(t)+3(t-4\M^2)(t-2\M^2)(1+C_{tu})
{1\over\pi}\intlow{dx\over x^2} C_{st}\Im \W^E_-(x)\nonumber\\
&&b_1(t)=a_1(t)-3W_1^E(t)-3(t-2\M^2){1\over\pi}\intlow
{dx\over x^2(x+t-4\M^2)}\Im f_1^1(x)\nonumber\\
&&\hskip 5 cm
-(t-4\M^2){1\over\pi}\inthi {dx\over x^2(x-4\M^2)}\Im F^{(I_t=1)}(x,t)\ .
\ena
We have introduced the amplitudes with given isospin in the $t$-channel
which are defined, as usual, by
\be
F^{(I_t=a)}(s,t) = \sum_{I=0}^{2} (C_{st})_{aI} F^I(s,t) \ .
\en
These explicit relations between the old and the new subtraction
functions will only be useful at a later stage for the derivation
of  slowly convergent sum rules for the parameters $\alpha$ and
$\beta$. $\Z^E(s,t)$
has now the following expression:
\be\lbl{ZE}
\Z^E(s,t)=C_{st}\Big\{
\b_+(t)+(s-u)\b_-(t)+us[\h_+(t,us)+(s-u)\h_-(t,us)]\Big\}\ ,
\en
where the vectors $\h_+(t,us)$ and $\h_-(t,us)$ are formed
in analogy to $\a_\pm,\ \W_\pm$
(see \rf{apam}, \rf{Wpe}) from the
three functions
\be
h_{a}(t,us)={-1\over\pi}\inthi
{dx\,(2x+t-4\M^2)\,\Im F^{(I_t=a)}(x,t)\over
x(x+t-4\M^2)[x^2+x(t-4\M^2)+us]}\ ,\  a=0,2
\en
and
\be
h_1(t,us)={-1\over\pi}\inthi
{dx\,\Im F^{(I_t=1)}(x,t)\over x(x+t-4\M^2)[x^2+x(t-4\M^2)+us]}\ .
\en
As announced, these dispersive integrals extend over the 
high energy range $[E^2,\infty]$.
 
Up to this point, $E$ was bounded only from above by the requirement 
that the contribution from the D-waves can be neglected 
for energies smaller than $E$.
We can now also take $E$ to be sufficiently large compared
to twice the pion mass. In other
words we require
\be
4\M^2<< E^2 < 1 \ {\rm GeV}^2\ .
\lbl{Erange0}
\en
Then, in the low energy region, the ratios $s/E^2,\ t/E^2,\ u/E^2$
will be of order  $O(p^2)$ in the chiral counting; consequently,
we can expand the integrals
$h_a(t,us)$ in powers of these quantities and
drop the terms which are of chiral
order $O(p^8)$ or higher in the amplitude 
(this is legitimate since we intend to
equate this form of the amplitude with the expansion to two-loop
chiral order). Hence, we can replace in Eq.\rf{ZE}
\be\lbl{hidev}
h_a(t,us)=h_a(0,0)+t\partial_t h_a(t,0)\vert_{t=0}+ O(p^4)\ (a=0,2),
\quad h_1(t,us)=h_1(0,0)+O(p^2)\ .
\en
Finally, we must impose $s-t$ crossing symmetry (to the same
chiral order) on $\Z^E(s,t)$, 
{\it i.e.}
\be\lbl{cross}
\Z^E(t,s)=C_{st}\,\Z^E(s,t) +O(p^8)\ .
\en
The standard trick of setting $s=0$ in \rf{cross} shows that the 
subtraction functions $b_a(t)$ must be polynomials 
once the integrals $h_a(t,us)$ 
have been expanded \rf{hidev}. These polynomials are 
entirely determined up to 
two constants, $a$ and ${b}$:
\bea
&& b_0(t)={{\M^2}\over 3}(5 a - 8b )
+2tb-{1\over 3}t(t-4\M^2)\left[ h_0+6(t-2\M^2)h_1+
5 h_2\right]\nonumber           \\
&&\nonumber \\
&& b_1(t)={b} -{t\over 6}\,\left
[2h_0+6(t-2\M^2)h_1-5 h_2\right]     \lbl{ba}    \\
&&\nonumber \\
&& b_2(t)={{2\M^2}\over 3}({a} +2 {b}) 
-tb-{t\over 6}(t-4\M^2)\left
[2 h_0-6(t-2\M^2)h_1+ h_2\right]\ ,\nonumber 
\ena
where we have introduced the notation
\be\lbl{ha}
h_a\equiv h_a(0,0),\qquad h'_a\equiv \partial_t h_a(t,0)\vert_{t=0}\ .
\en
The crossing relation \rf{cross} is then identically satisfied
provided the following combination
of high energy integrals  vanishes in the chiral limit:
\be
2h'_0-5h'_2-9h_1= O(p^2)\ .
\en
(Using resonance saturation, this constraint
may be converted into an amusing relation between the
masses and the widths of the $f_2(1270)$ and the
$\rho_3(1690)$ resonances
\footnote{
This relation reads: $63\Gamma_\rho M_f^4=5\Gamma_f M_\rho^4$, where
$M_f$ and $M_\rho$ are the masses of the $f_2$ and the $\rho_3$ mesons
and $\Gamma_f,\ \Gamma_\rho$ are the respective \pp partial widths.
}.)
This completes the construction of a dispersive representation 
of the amplitude $\F(s,t)$ in terms of two arbitrary
parameters, ${a}$ and ${b}$. 
Explicitly, from \rf{WE} and \rf{ba}  one obtains the following expression
for the corresponding function $A(s\vert t,u)$
\be\lbl{Adisp}
\begin{array}{rr}
{1\over{32\pi}}\,A^{disp}(s\vert t,u)=
{1\over3}\left[ W^E_0(s)-W^E_2(s)\right] +{3\over2}(s-u)W^E_1(t)
+{1\over2}W^E_2(t)&\ \\
+{3\over2}(s-t)W^E_1(u)+{1\over2}W^E_2(u)&\ \\
+h_1(-s^3+6\M^2s^2-8\M^4s)+{1\over3}(h'_0-h'_2)stu-{1\over2}h_2
s(s-4\M^2)+{1\over3}(h_0-h_2)tu& \ \\
+{b} s+ {1\over3}({a}-4{b})\M^2&  +O(p^8)\ . \\
\end{array}
\en

Expression \rf{Adisp} represents the final form of the dispersive
\pp amplitude.
We recall that the functions $W_a^E$ are 
computed from the imaginary parts of the partial-wave amplitudes
$f_0^0(s)$, $f_1^1(s)$ and $f_0^2(s)$ (see \rf{Wae})
in the domain $4\M^2\le s\le E^2$,
{\it i.e.} in the region where experimental information is incomplete, while
the constants $h_a$ and $h'_a$ \rf{ha}
can be evaluated using existing 
experimental data at higher energy: the part involving these functions
corresponds to the so-called driving terms in the 
Roy equations \cite{bfp}. 
Indeed, from $A^{disp}(s\vert t,u)$ one can
compute the real parts of $f_0^0$, $f_1^1$ and $f_0^2$, which are thus
given in terms of the corresponding 
imaginary parts and the driving terms. These
integral equations, together with the elastic unitarity relations (which
hold to good accuracy below the $K\bar K$ threshold) constitute the
Roy equations. Numerical solutions to these equations were constructed
in \cite{bfp}\cite{pennington} taking as the two free parameters 
the scattering lengths $a_0^0$ and $a_0^2$.
Constraining these solutions to match with the experimental data in the 
$600-800$ MeV region, it was shown that 
$a_0^0$ is left essentially undetermined but that $a_0^2$ gets strongly
correlated with $a_0^0$   
(this relationship is known as the Morgan-Shaw band in the literature
\cite{mmsbook}). 
We are going to show that in a large domain of parameters $\alpha$
and  $\beta$, the perturbative amplitude $A^{2-loop}(s\vert t,u)$ 
of Eq.\rf{A1} provides a rather accurate representation of the numerical 
solutions of the Roy equations (up to energies $\sqrt{s}\approx
0.5$ GeV), provided the values of the parameters $\lambda_1$, $\lambda_2$, 
$\lambda_3$ and $\lambda_4$ are properly chosen.

\bigskip
\noindent{\bf 2.2 Equating the dispersive and perturbative amplitudes}

\bigskip
Let us now restrict ourselves to the
domain where the Mandelstam variables $s,\ t,\ u$ are very small
(compared to 1 GeV$^2$). In this domain,  
$A^{disp}(s\vert t,u)$ must be identical with 
$A^{2-loop}(s\vert t,u)$, except for terms
of chiral order $O(p^8)$ or higher. 
Let us then subtract the part which contains
all the cuts of the two-loop expression, $A^{cut}(s\vert t,u)$ 
(see \rf{Acut}) 
from the dispersive representation \rf{Adisp} of $A(s\vert t,u)$: 
this difference must be identical to the polynomial part of the
chiral amplitude, $A^{pol}(s\vert t,u)$ (c.f. Eq.\rf{Apol}). 
Indeed, at low energies, the discontinuity in the differences
\be
\Im \left[ W^E_a(z)-\WW_a(z)\right]= \Im f^a_0(z)-\Im f^a_0(z)\vert_{2-loop}\ 
\quad (a=0,2)
\en
is, by definition, a quantity of chiral order eight. 
A similar relation also holds for the isospin one case.
In the range where
$z<<1$ GeV$^2$, these  discontinuities can be neglected and the differences
$W^E_a(z)-\WW_a(z)$
are analytic functions which are well-approximated by polynomials
\be
32\pi[W_a^E(z)-\WW_a(z)]=\sum_{k=0}^3 I_a^k z^k + O(z^4)\ .
\lbl{Iak}
\en
Inserting these expansions into the difference
$A^{disp}(s\vert t,u)-A^{2-loop}(s\vert t,u)$, 
we generate six sum rules by recognizing that
the polynomial obtained in this way should vanish up to terms of order
$O(p^8)$. 
It is then a simple exercise to derive:
\bea
&&{\displaystyle\lambda_1\over\displaystyle \fpi^4}
={1\over 3}(I_0^2-I_2^2) -3I_1^1+2\M^2(I_0^3-I_2^3-3I_1^2)
+{{16\pi}\over3}(h_0-4h_2)\nonumber\\
&&\nonumber \\
&&{\displaystyle \lambda_2\over\displaystyle \fpi^4}
={1\over 2} I_2^2+{3\over 2} I_1^1
+3\M^2(I_2^3+I_1^2) -{{16\pi}\over 3}(h_0-h_2)\nonumber\\
&& \lbl{sumrule} \\
&&{\displaystyle \lambda_3\over\displaystyle \fpi^6}
={1\over 3}(I_0^3-I_2^3)
+I_1^2+{{32\pi}\over 9}(h'_0-h'_2-9h_1)\nonumber\\
&&\nonumber \\
&&{\displaystyle \lambda_4\over\displaystyle \fpi^6}
={1\over 2} (I_2^3- I_1^2) +{{32\pi}\over 9}(h'_0-h'_2)\ ,\nonumber
\ena
where the  entries $I_a^k$ are defined in \rf{Iak}.

The remaining two equations similar to \rf{sumrule} concern
the parameters $\alpha$ and $\beta$. However, these
depend explicitly on the  two subtraction constants ${a}$ 
and ${b}$ of the Roy representation \rf{Adisp}: 
\bea
&&{\displaystyle \alpha\over\displaystyle\fpi^2}=
             {4\over 3}\left[ I_0^1+2I_2^1 +\M^2(I_0^2+2I_2^2)\right]+
32\pi\left[2\M^2(h_0+2h_2)+{8\over3}\M^4(h'_0-h'_2-3h_1)
+{a}\right]\nonumber\\
&& \lbl{betasum0}  \\
&&{\displaystyle \beta\over\displaystyle \fpi^2}=
               {1\over 3} I_0^1-{5\over 6} I_2^1     + 4\M^2
              \left( {1\over 3} I_0^2-{5\over 6} I_2^2\right)+12\M^4
              \left({1\over 3} I_0^3-{5\over 6} I_2^3-{1\over 2} I_1^2\right)
+32\pi\left[ 4\M^4 h_1 +{b}\right]\ .\nonumber
\ena
Consequently, one cannot, {\it a priori}, make use of these equations
to determine $\alpha$ and $\beta$. 

The sum rules \rf{sumrule} and \rf{betasum0} represent the minimal
set of necessary and sufficient conditions ensuring the compatibility
of the dispersive representation \rf{Adisp} with the chiral expansion
of the amplitude up to and including the two-loop accuracy. It is worth
stressing the role of crossing symmetry in the derivation of these 
sum rules. 

\bigskip
\noindent{\bf 2.3 Barely converging sum rules for the parameters
$\alpha$ and $\beta$}

\bigskip
The convergence of the high-energy integrals in Eqs. \rf{sumrule}
is guaranteed by the Froissart bound. Actually, as will be seen in
sec. 3, the sum rules \rf{sumrule} are rapidly convergent and rather
independent of the details of high-energy asymptotics. Additional
information on the high-energy behaviour of the amplitude would yield
additional sum rules. The standard Regge pole phenomenology, 
for instance, suggests that the $I_t=1$ $t$-channel isospin amplitude
$F^{(I_t=1)}(s,t)$ behaves asymptotically as $s^{\alpha_\rho(t)}$ where
$\alpha_\rho(t)$ is the $\rho$ trajectory with the intercept
$\alpha_\rho(0)\simeq 1/2$. According to this picture, the amplitude
$F^{(I_t=1)}$ satisfies a once-subtracted dispersion relation. This idea 
was used in the past to generate a sum rule for the P-wave scattering
length $a_1^1$ and for the 
combination $2a_0^0-5a_0^2$ \cite{olsson}. Comparing the 
once-subtracted with the twice-subtracted fixed t dispersion 
relation \rf{disp}, one obtains the following expression for the 
subtraction function $a_1(t)$:
\be\lbl{a1}
a_1(t)={1\over\pi}\int_{4\M^2}^\infty {dx\over x^2}\,\Im
F^{(I_t=1)}(x,t)\ .
\en
Similarly, the amplitude with $F^{(I_t=2)}(s,t)$ is dominated by exotic 
$t$-channel exchanges and it is expected to be asymtotically suppressed. 
It is even conceivable that it satisfies an unsubtracted dispersion
relation \cite{mmsbook}. Under this assumption, one obtains the 
following sum rule for the subtraction function $a_2(t)$ in \rf{disp}:
\be\lbl{a2}
a_2(t)={1\over\pi}\int_{4\M^2}^\infty {dx\over x^2}\,
(2x+4\M^2-t)\Im F^{(I_t=2)}(x,t)\ .
\en
The integral \rf{a2}, if convergent at all, converges even more slowly
than the sum rule \rf{a1}. The asymptotic behavior of $F^{(I_t=2)}$ is
expected to be dominated by $s^{2\alpha_\rho-1}$ arising from the 
$\rho-\rho$ Regge cut \cite{regge}. Assuming the validity of Eqs. \rf{a1}
and \rf{a2} at $t=0$, one can use them in \rf{subrel} and then, from
\rf{ba} express the two subtraction constants $a$ and $b$:
\bea
b=&&{1\over6\pi}\intlow {dx\over x^2} \Im \Big\{ 2f_0^0(x)-5f_0^2(x)+
{9x\over x-4\M^2}f_1^1(x)\Big\} \nonumber\\
&&+{1\over\pi}\inthi{dx\over x(x-4\M^2)}\Im F^{(I_t=1)}(x,0) ,\lbl{bsum}\\
(a+2b)\M^2=&&{1\over2\pi}\intlow {dx\over x}\Im\Big\{
(1+{2\M^2\over x})[2f_0^0(x)+f_0^2(x)]-9{x-2\M^2
\over x-4\M^2}f_1^1(x)\Big\}\nonumber\\
&&+{3\over\pi}\inthi{dx\over x^2}(x+2\M^2)\Im F^{(I_t=2)}(x,0) . \lbl{asum}
\ena
These relations promote Eqs. \rf{betasum0} to two  additional sum rules
for $\alpha$ and $\beta$. 

The slow convergence of the high-energy integrals in Eqs. \rf{bsum} and
\rf{asum} 
prohibits any practical use of these sum rules for $\alpha$ and $\beta$. 
The sum rule for $\beta$, which is expected to converge better, leads
to values $\beta=1.2-1.4$ which have the correct order of magnitude (as
compared to the {\chpt} prediction), but the uncertainty due to the 
high-energy tail of the integral in \rf{bsum} is difficult to estimate. 
Needless to say, the evaluation of the sum rule for $\alpha$ 
which involves the integral \rf{asum} is even more problematic. 

In the following, these barely convergent sum rules for $\beta$ and $\alpha$
will be ignored. The above discussion should mainly serve as an explanation
of why, in contrast to $\lambda_1$,...,$\lambda_4$, the two parameters 
$\alpha$ and $\beta$ cannot be determined from the existing \pp scattering
data at medium energies. New, high precision low-energy experiments are, 
indeed, unavoidable for this purpose.

\bigskip
\noindent{\large\bf 3. Evaluation of the sum rules}

\bigskip
In this section, the details of the evaluation of the four sum rules
\rf{sumrule} are presented. The input data above 0.5 GeV and their use are
discussed in subsection 3.1. The treatment of the low energy part is explained
in subsection 3.2, where the final results are also presented. Subsection 3.3
is devoted to a careful analysis of error bars. Finally, subsections 3.4 and
3.5 contain a comparison with related works, in particular with the recently
published standard $\chi$PT two loop calculation \cite{bij}.

\bigskip
\noindent{\bf 3.1 Energy region above 0.5 GeV}

\bigskip
The main sources of experimental information concerning
the \pp amplitude are the production experiments (see \cite{mmsbook}),
which give reliable results in the range $0.5\le\sqrt{s}\le 1.9$ GeV.
These, together with \rf{Erange0}, 
imply that the terms $h_i$ and  $h'_i$, which are integrals over
the range $[E^2,\infty]$,  can essentially be evaluated using 
experimental data.  The contribution from the asymptotic
region $\sqrt{s}>1.9$ GeV can be estimated using
Regge phenomenology and it turns out that this contribution is
fairly small. This is illustrated in table 1 below, which shows the
contributions of the various energy ranges to these integrals.
\begin{center}
\begin{tabular}{|c|c|c|c|c|} \hline
$\sqrt{x}$\,GeV & [0.5,0.95] &  [0.95,1.9] & $[1.9,\infty]$ & total\\ \hline
$h_0(\GeV^{-4})$ & -6.15     &   -0.59     &     -0.08      &-6.82 \\ \hline
$h_2(\GeV^{-4})$ &  0.17     &   -0.15     &      0.00      & 0.02 \\ \hline
$h_1(\GeV^{-6})$ & -4.34     &   -0.11     &     -0.002     &-4.45 \\ \hline
$h'_0(\GeV^{-6})$ & 8.15  &   -0.54     &     -0.13      & 7.48 \\ \hline
$h'_2(\GeV^{-6})$ &11.82 &   -0.20     &      0.00      &11.62 \\ \hline
\end{tabular}
\vskip 0.5 true cm
{\bf Table 1:\ }{\sl Contributions from the various energy ranges to the
integrals $h_i$ and $h'_i$, with $E=0.5$ GeV}
\end{center}
The table shows clearly that the integrals are rapidly convergent
and are dominated by the region $\sqrt{s}<1$ GeV with the exception,
however, of $h_2$. In that case, the integrand is very
small below 1 GeV because of cancellations among
the three isospin contributions, and the whole integral is
much suppressed as compared to the others.
This is reminiscent of the
approximation made in \cite{pp} of setting the combination of
amplitudes with $I_t=2$ equal to zero in sum rules 
derived there for $\lambda_1$,
$\lambda_2$. One should be aware that this simple approximation breaks
down for $h'_2$;  as can be seen from the table, $h'_2$ is even larger
than $h'_0$.
 
We now specify in more detail how we have treated the experimental
data in order to obtain the numbers given above and the central values
for the $\lambda_i$'s as given in the sequel. Let us first discuss the 
region below 1.9 GeV. The region was divided into two subregions 
which were treated
somewhat differently: 
$$
a)\, 0.95\le\sqrt{s}\le1.9 \ {\rm GeV}\quad {\rm and}\quad 
b)\, 0.5\le\sqrt{s}\le0.95 \ {\rm GeV} 
$$
\begin{itemize}
\item
In the subregion a) we have employed the data 
obtained by Hyams {\it et al.} \cite{hyams} (based on the 
CERN-Munich production experiment \cite{grayer}, which has by far 
the best statistics to date) in the form of the
analytic K-matrix type parametrization which the authors provide \footnote{
There are a few obvious misprints in the published formulae.}. 
Phase shifts and inelasticities have been determined for partial
waves up to $\ell=3$, which means that the contributions from the 
resonances $f_2(1270)$ and $\rho_3(1690)$ are automatically included. 
Partial waves with $\ell\ge 4$ were ignored: including the resonance 
$f_4(2040)$ in the narrow width approximation proved to have a
negligibly small effect.
In the case
of the isospin $I=2$  and the partial wave $\ell=0$, we
have used the data of  Hoogland {\it et al.}  
\cite{hoogland} which have better statistics in this channel,
and we have neglected
the contributions of the higher partial waves. 
\item
In the subregion b), some care is needed 
because the results are very
sensitive to the values of the phases there, particularly to the S
and P partial waves.
Concerning the $I=0$ S-wave, several
slightly different analyses of the CERN-Munich data have been performed. 
We shall use, as before,  the data of Hyams {\it et al.} which extend
down to 0.6 GeV, and also the results of Estabrooks and Martin
\cite{em} who provide the
phases down to 0.5 GeV. We have performed a fit to both sets of data
using the  parametrization proposed by Schenk \cite{schenk}:
\be\lbl{schenk}
\tan\delta_\ell^I(s)= p^{2\ell}\sqrt{1-{4\M^2\over s} }
\left({4\M^2-s_0\over s-s_0}\right)
(a_\ell^I + c p^2 +d p^4)\ ,\qquad p^2={s\over4\M^2}-1\ ,
\en 
where 
$c$, $d$ and $s_0$ are three parameters to be determined by the fit. 
In principle, one could take $a_\ell^I$ to be a free parameter as well, 
since we intend to use \rf{schenk} rather far from the threshold, in
the energy region from 0.5 to 1 GeV. However, we have preferred to keep
with the  idea of \cite{schenk}, taking
$a_\ell^I$ to be the experimental value of the 
scattering length, in order not to give too much weight to the first few
points around 0.5 GeV.
The quality of the fit is practically independent of the
exact value of $a_\ell^I$, within the range allowed by experiment.
The numerical
values that we have used are collected in table 2. In the case of the 
P-wave, the data deduced from the CERN-Munich
experiment are known to be in slight conflict 
with the results based on the 
Roy equations 
below 0.7 GeV \cite{manner,basdevant}. In that case, 
the Roy equation studies suggest that
the $\delta_1^1$ phase is very close to a pure Breit-Wigner shape below
the resonance. Based on this prejudice, we have again used the
parametrization \rf{schenk},  
imposing the experimental value of the scattering length and 
adjusting the remaining
three parameters in order to fit the experimental values of the
$\rho$ mass and of the $\rho$ width, and such that $\delta_1^1(\sqrt{s}=
0.95$ GeV) matches to the experimental value of \cite{hyams}. If one
uses a real fit to the data of \cite{hyams} instead, then the 
difference in the results would be of the order of 5\%, which
means that they are 
perfectly compatible within the errors.
Finally, for the $I=2$ $\ell=0$
partial wave we have fitted the parameters of the representation
\rf{schenk} to the data of \cite{hoogland}. In that case, we have used the
same parametrization in both region a) and region b).
\end{itemize}
For comparison, we have also used the parametrizations of the $I=0$ S-wave
given by \cite{au,krupa}, of the $I=1$ P-wave given by \cite{erkal} and the
$I=2$ S-wave given by \cite{hyams}. Our final results are consistent with these
fits to the data.

\medskip
\begin{center}
\begin{tabular}{|c|c|c|c|c|}\hline
\          &$a_\ell^I$ & $c$  &  $d$    &  $\sqrt{s_0}\,\rm{(MeV)}$ \\ \hline
I=0\ $\ell=0$ & 0.26   & 0.2535         &$-0.0200$         &$843$\\ \hline
I=2\ $\ell=0$ &$-0.028$  &$-0.2325$       &$-0.0160$         &$0$\\ \hline
I=1\ $\ell=1$ & 0.038    &$ 0.2560\,10^{-3}$&$-0.6009\,10^{-4}$&$769$\\ \hline
\end{tabular}
\vskip 0.5 truecm
{\bf Table 2:\ }{\sl\ Numerical values of the parameters  used 
in \rf{schenk} for the S and P partial waves. The values of the
scattering lengths are taken from \cite{dumbrajs} }
\end{center}

In the asymptotic region we have assumed that the amplitudes are 
given by Regge phenomenology, and we have used the same parameters as in
formulae (17), (18) and (19) of \cite{fp}. This representation was 
assumed to hold for $\sqrt{s}\ge3$ GeV, and in the region between
1.9 and 3 GeV the amplitude was determined by linear interpolation.

\bigskip
\noindent{\bf 3.2 Energy region below 0.5 GeV and results}

\bigskip 
We must now deal with 
the terms $I_a^k$ in the sum rules.
According to the definition \rf{Iak},
they are obtained by making a Taylor expansion of 
the difference $W_a^E(z)-\WW_a(z)$. Here, $\WW_a(z)$ is known analytically
in terms of $\alpha$, $\beta$ and is a linear function of $\lambda_1$,
$\lambda_2$. Since $E$ is constrained not to exceed 1 GeV, one can
ignore inelasticities to a good approximation and, using unitarity, 
express $W_a^E$ in terms of phase shifts:
\bea\lbl{Wsin}
&&W_{a}^E(z)={s^2\over\pi}\intlow{dx} \sqrt{{x\over x-4\M^2}}
{\sin^2\delta_0^a(x)\over x^2(x-z)}\quad (a=0,2)\ ,\nonumber\\
&&W_{1}^E(z)={s\over\pi}\intlow{dx}\sqrt{{x\over x-4\M^2}}
{\sin^2\delta_1^1(x)\over x(x-4\M^2)(x-z)}\ .
\ena
In the upper part of the
integration range, $0.5\ {\rm GeV}\le\sqrt{x}\le E$, 
we can use the experimental
phase shifts as discussed above. 
Below 0.5 GeV, experimental results for the phase shifts are rather
limited. Fortunately, it is precisely in this range that we can make use
of the chiral expansion of the amplitude. One expects that the chiral
two-loop representation for  the
three phase shifts $\delta_0^0,\ \delta_0^2$ and $\delta_1^1$ should be
reasonably precise in the range $2\M \le\sqrt{s}\le500$ MeV: this 
expectation will be verified more quantitatively by comparing our 
results with those
of a numerical solution of the Roy equations. We will 
therefore employ the chiral representation of the phase shifts 
in the dispersive integrals \rf{Wsin} below 0.5 GeV.
When calculated
in this way, the set of parameters $I_a^k$ depend on $\alpha$, $\beta$ and
also, in a nonlinear way, on the four $\lambda_i$'s.
A less precise procedure, but equally acceptable to the desired
chiral order,
is to use, below 500 MeV, the imaginary parts of the S and P
partial wave amplitudes 
in the chiral two-loop approximation (which is known to
violate unitarity appreciably close to 500 MeV); {\it i.e.}, 
one uses in the integrands of Eq.\rf{Wae}
\bea\lbl{Wapprox}
&&\Im f_0^{a}(x)=\sum_i w_a^i(x)\Im\bar K_i(x)\ \quad (a=0,2)\nonumber\\
&&\Im f_1^1(x)=(x-4\M^2)\sum_i w_1^i(x)\,\Im\bar K_i(x)\ .
\ena
In this approximation, the
dependence on $\lambda_1$, $\lambda_2$ is simply linear and there is
no dependence upon $\lambda_3$ and $\lambda_4$.
In general, the  system of
equations \rf{sumrule} must be solved in a self-consistent way.
 
Let us now discuss the numerical results. First, we consider the
stability with respect to variation of the energy parameter $E$. 
Obviously, the parameters $\lambda_1,...,\lambda_4$ should be
independent of $E$. The validity of the formulas \rf{sumrule}, 
as discussed in the preceeding section, requires that $E$ should
be smaller than, roughly, 1 GeV  (see the discussion preceeding \rf{Imf})
and at the same time that
$E^2>>4\M^2$. These conditions suggest to consider
an interval
\be
500\ {\rm MeV} <E< 1000\ {\rm MeV}\ .
\en 
Variations of the results within this interval 
are shown in table 3 below \footnote{As in I, we use $F_{\pi}$ = 92.4 MeV and
$M_{\pi}\,=\, M_{\pi^+}\,=\,$139.57 MeV.}
 
\begin{center}
\begin{tabular}{|l|rrrrr|}\hline
E (MeV)        & 500\ & 600\ & 700\ & 800\ & 900\ \\ \hline
$10^3\lambda_1$&-4.86 &-5.82 &-6.04 &-6.08 &-6.11 \\
$10^3\lambda_2$& 9.68 & 9.64 & 9.60 & 9.56 & 9.55 \\
$10^4\lambda_3$& 3.33 & 2.55 & 2.31 & 2.20 & 2.16 \\
$10^4\lambda_4$&-1.46 &-1.49 &-1.48 &-1.46 &-1.45 \\  \hline
\end{tabular}
\vskip 0.5 true cm
{\bf Table 3:\ }{\sl Results from the sum rules \rf{sumrule} for several
values of the energy parameter $E$. The numbers correspond to
$\alpha=2$ and $\beta=1.08$}
\end{center}
Clearly, one does not observe exact stability. The variation is
more significant for $\lambda_1$ and $\lambda_3$ than for the remaining
two parameters, which are stable within 5\%. A reasonable stability 
plateau forms for all four parameters for values of $E$ above $700-800$
MeV. Keeping in mind that $E$ should remain sufficiently small
such that the imaginary part of the D-wave contributions can be
neglected,
we shall assume in what follows that $E=800$ MeV provides a reasonable 
compromise.

Let us now discuss the results for the $\lambda_i$'s 
corresponding to various
values of the two  parameters $\alpha$ and $\beta$. 
This variation is of interest for the purpose of using the chiral
formulas in a fit to determine $\alpha$ and $\beta$ from experiment. 
Recall 
that $\alpha$ is expected to be close to 1
(within, say, 20\%) according to standard chiral
perturbation theory. In contrast, the generalized \chpt\  can accomodate 
values of 
$\alpha$ as large as $\alpha=4$.  At present, the best
$K_{l4}$ data available \cite{rosselet} are compatible with a relatively
wide range of values, $1\lapprox\alpha\lapprox3$. 
Some results, illustrating the dependence on $\alpha$ 
and $\beta$, are collected in table 4
below (the numbers shown correspond to $E=800$ MeV).

\medskip
\begin{center}
\begin{tabular}{|c|c||c|c|c|c|} \hline
$\alpha$  & $\beta$ &
  $10^3\lambda_1$ & $10^3\lambda_2$ &$10^4\lambda_3$ &$10^4\lambda_4$\\ \hline
1.04  & 1.08 & -5.68 & 9.32 & 2.21 & -1.48 \\ \hline\hline
2.00  & 1.00 & -6.60 & 9.39 & 2.12 & -1.33 \\
2.00  & 1.08 & -6.08 & 9.56 & 2.20 & -1.46 \\
2.00  & 1.13 & -5.70 & 9.67 & 2.26 & -1.54 \\ \hline\hline
2.50  & 1.08 & -6.27 & 9.68 & 2.19 & -1.45 \\ \hline
3.00  & 1.08 & -6.45 & 9.78 & 2.19 & -1.44 \\ \hline
3.50  & 1.08 & -6.62 & 9.88 & 2.18 & -1.44 \\ \hline
4.00  & 1.08 & -6.79 & 9.97 & 2.17 & -1.43 \\ \hline
\end{tabular}
\vskip 0.5 true cm
{\bf Table 4:\ }{\sl Results for the parameters 
$\lambda_1$,...,$\lambda_4$ corresponding
to several values of $\alpha$ and $\beta$.}
\end{center}
The first line in the table corresponds to the values of $\alpha$ and
$\beta$ obtained in the standard \chpt\ at chiral order $O(p^4)$ in the three
flavour case (see I).
For the particular value $\alpha=2$, we have varied $\beta$ in the
range allowed by the Morgan-Shaw band. 
The table shows that as long as 
$\alpha$ and $\beta$  remain in the range allowed by experiment, the
variation of the values of the $\lambda_i$'s remains smaller than
the error bars (see table 6). As a first approximation one may
ignore this variation and adopt as average values
\be
10^3\lambda_1\simeq-6.1,\quad
10^3\lambda_2\simeq9.6,\quad 10^4\lambda_3\simeq2.2,\quad
10^4\lambda_4\simeq-1.45\ .
\en
These numbers update those given in our previous paper I
\be
10^3\lambda_1\simeq-5.3,\quad
10^3\lambda_2\simeq9.7,\quad 10^4\lambda_3\simeq2.9,\quad
10^4\lambda_4\simeq-1.40\ ,
\en
which were computed using a smaller value of $E$ ($E=550$ MeV),
and with a slightly different treatment of the experimental data.
The results are compatible within the uncertainties and
the differences do not affect 
the numbers given in I for the threshold parameters. The results at higher
energies are obviously more affected by small modifications of the
values of $\lambda_1$,...$\lambda_4$. For instance, computing
the value of the phase of the $\epsilon'$ parameter  with the numbers 
from table 4, one would find a value higher by $2-3^\circ$ than 
the one given in I.

Once the $\lambda_i$'s have been determined self-consistently, one can
return to the sum rule expressions and examine the contribution of each
integration region. These are displayed in table 5 which shows that
the contribution from the resonance region, 0.5 GeV to 1 GeV, dominates the
values of the three parameters $\lambda_1,\ \lambda_2$ and $\lambda_3$,
while the contribution from the low energy region is smaller though
not negligible. The situation is different in the case of $\lambda_4$,
for which the low energy region dominates. One may then fear that 
$\lambda_4$ could be overly sensitive to higher order chiral corrections. 
However, this does not seem to be the case: an estimate of the $O(p^8)$
uncertainties is performed in the next subsection which shows that
$\lambda_4$ is weakly affected (see table 6).

\begin{center}
\begin{tabular}{|c|c|c|c|c|}\hline
$\sqrt{x}$ (GeV) &[$2\M$,0.5] & [0.5,1] & [1,1.9]&[1.9,$\infty$]\\ \hline
$10^3\,\lambda_1$& 2.72       &-8.84    & 0.14   &-0.10       \\
$10^3\,\lambda_2$& 1.35       & 7.69    & 0.42   & 0.10        \\
$10^4\,\lambda_3$& 0.49       & 1.68    & 0.03   &-0.001        \\
$10^4\,\lambda_4$&-1.15       &-0.28    &-0.02   &-0.001        \\ \hline 
\end{tabular}
\vskip0.5 truecm
{\bf Table 5:\ }{\sl Contributions of four successive integration ranges 
to the values of $\lambda_1$,...$\lambda_4$ with $\alpha=2$, 
$\beta=1.08$.}
\end{center}

\newpage
\bigskip
\noindent{\bf 3.3 Discussion of the errors}

\bigskip
We can identify two sources of uncertainties in the estimate 
of the $\lambda_i$'s: i) the errors
affecting the experimental phase shifts and ii) the errors affecting the
theoretical phase shifts, {\it i.e.} contributions of chiral order $O(p^8)$
or higher that one might expect to become sizable in the neighborhood
of 500 MeV. In order to estimate the uncertainties arising from the former
source, we proceed as follows.
First we take $E$ to be $E=0.5$ GeV.  
Using this value for $E$ has the advantage that the experimental
phases show up only in the integrals $h_i$, $h'_i$ and not in the
terms $I_a^k$. In computing $I_a^k$, we may use the
approximation \rf{Wapprox} so that the sum rules \rf{sumrule} can be
solved analytically (since they reduce to a linear system) for
$\lambda_1$,...,$\lambda_4$.  
Finally, we  assume that the errors on the
phase shifts are piecewise constant
as a function of energy
(this assumption is seriously violated only in a narrow region
around the $K\bar K$ threshold), 
and we have adopted the following values, an educated guess inspired
by \cite{hyams,em,hoogland}:
\be
\begin{array}{llll}
0.50\le\sqrt{s}\le0.95{\rm GeV}&\quad\Delta\delta_0^0=4^\circ
&\quad\Delta\delta_1^1=1^\circ&\quad\Delta\delta_0^2=2^\circ\\
0.95\le\sqrt{s}\le1.90         &\quad\Delta\delta_0^0=9^\circ
&\quad\Delta\delta_1^1=1^\circ&\quad\Delta\delta_0^2=4^\circ\ .\\
\end{array}
\en
Under these assumptions the errors are given as simple analytic 
expressions in terms of a few phase shift integrals and it is a
simple matter to calculate them. 
With the choice $E=0.5$ GeV, table 1 shows that
most of the contributions
to the integrals $h_i$ are  concentrated in the integration
region below 1 GeV. We have therefore neglected the errors coming
from either the inelasticities or the partial waves with $\ell\ge2$, or
from the asymptotic contributions.
Finally, in order to estimate the error coming from the missing higher
chiral orders in the low energy integrals, we have compared the result
of calculating $W_a^E(z)$ using three evaluations of the integrands
which have identical chiral expansions up to (and including) chiral order six, 
and which differ at higher chiral order:
~i) we use expressions \rf{Wsin}
together with the chiral approximation to the phase shifts
(see formula (4.13) of I )  
~ii) we use the same expression but get the phase shift from a 
Pad\'e approximant to the pertubative expansion which 
satisfies unitarity exactly (this
generalizes to the $O(p^6)$ case  
\footnote{ For the S and P partial wave amplitudes,
the chiral expansion consists
in three successive terms $f=f^{(2)}+f^{(4)}+f^{(6)}$ and one finds
$f^{Pad\acute e}=f^{(2)}/\big[1-f^{(4)}/f^{(2)} +(f^{(4)}/f^{(2)})^2 -
f^{(6)}/f^{(2)}\big] $. }
the idea used in the context of \chpt\ 
in  \cite{dobado}) 
and ~iii) we use the
chiral approximation \rf{Wapprox}, which violates unitarity. We have
collected the results of these error estimates in table 6 where we
show separately the contribution from each source, as discussed 
above.

\begin{center}
\begin{tabular}{|c|c|c|c|c||c|}\hline
\ &$\Delta\delta_0^0$ &
   $\Delta\delta_1^1$ &
   $\Delta\delta_0^2$ &
                                     $O(p^8)$  & Total \\ \hline
$10^3\Delta\lambda_1$&1.02 &1.47 & 0.40 & 0.40 & 2.23 \\ \hline
$10^3\Delta\lambda_2$&0.05 &0.26 & 0.18 & 0.20 & 0.52 \\ \hline
$10^4\Delta\lambda_3$&0.34 &0.25 & 0.14 & 0.20 & 0.64 \\ \hline
$10^4\Delta\lambda_4$&0.01 &0.03 & 0.09 & 0.02 & 0.12 \\ \hline
\end{tabular}
\vskip 0.5 true cm
{\bf Table 6:\ }{\sl Errors on $\lambda_1$,...,$\lambda_4$ arising 
respectively from the experimental errors on $\delta_0^0$, $\delta_1^1$
and $\delta_0^2$ and from $O(p^8)$ corrections to the amplitude 
below 0.5 GeV.}
\end{center}

The entries in the table were evaluated for $\alpha=2$, $\beta=1.08$, 
but the errors show no significant variation with $\alpha$ or $\beta$. 
We have added the errors associated with the experimental errors
on the three phase shifts in quadrature, as it seems reasonable to
assume that they are independent. The last piece of the error was
added linearly. The reason why $\lambda_2$ and $\lambda_4$ have a
small relative error compared to $\lambda_1$ and $\lambda_3$ is that
the contribution from the isospin $I=0$ amplitude essentially drops
out. (There is only an indirect small contribution which comes from
the fact that the determination of 
$\lambda_2$ and $\lambda_3$ depends on $\lambda_1$). The values of the
parameters $\lambda_2$ and $\lambda_4$ are thus to a large extent
controlled by the $\rho$ resonance.

\bigskip
\noindent{\bf 3.4  Comparison with former results}

\bigskip
The parameters $\lambda_3$, $\lambda_4$, being of chiral order $O(p^6)$,
have not been discussed previously in the literature. 
At $O(p^4)$, a potentially accurate
determination of the parameters $\lambda_1$ and $\lambda_2$ 
is possible, making use
of the linear relationship with the low energy constants
$L_1$, $L_2$ and $L_3$  (see I: the relation is the
same in the standard and in the generalized \chpt). These three constants 
can be separately determined from the form factors of the $K_{l4}$ decay 
amplitude. On the theoretical side, this amplitude was analyzed at
order one loop of the \chpt\ \cite{bij1,rigg} and estimates of higher
order corrections were made \cite{bij2}. New high statistics data on
$K_{l4}$ decays would enable a  fairly
accurate determination of these constants. 
The best results available at present \cite{bij2}, which make use 
of the 
experimental results of Rosselet {\it et al.} \cite{rosselet} together
with the values of the \pp D-wave  scattering lengths quoted in
\cite{dumbrajs}, lead to
\be\lbl{lambgl}
\lambda_1=(-6.4\pm6.8)\,10^{-3}\ ,\qquad\lambda_2=(10.8\pm1.2)\,10^{-3}
\quad{\rm(one\ loop)}\ .
\en
If one uses solely the constraints from the D-wave scattering lengths, then,
according to \cite{gl84,bij2}, the error on $\lambda_1$ would be increased by
50\% and the error on $\lambda_2$ would be three times larger,
while the central values would remain approximately the same. 
In comparing these central values with the numbers that we quote in
table 3, one must keep in mind that in the latter $\lambda_1$ and
$\lambda_2$ include corrections of chiral order six. 
As far as the order of magnitude of
these corrections is concerned, 
a reasonable guess should be provided by the values of 
$\lambda_3$ or $\lambda_4$. This leads one to expect that these corrections
should not exceed $2-3\%$. On the
contrary, we find that the 
$O(p^6)$ corrections to the D-wave scattering lengths are significant. 
This is illustrated in table 7, where the values of a few 
threshold expansion
parameters which pick up their leading contribution at $O(p^4)$
are shown, both at $O(p^4)$ and at $O(p^6)$ (the values of 
$\alpha$, $\beta$, $\lambda_1$,...$\lambda_4$ being kept the same).

These contributions explain why our central value for $\lambda_2$ is 
smaller than \rf{lambgl} by roughly 10\%. Apart from this effect, we
believe that the errors on the LEC's (low energy coefficients)
extracted 
from these D-wave scattering lengths have been somewhat overestimated
by treating them as independent experimental data. In reality, the
numbers quoted in \cite{dumbrajs} are obtained as sum rules from the
Roy dispersive representation \rf{Adisp} projected on $\ell=2$, 
and are evaluated using experimental data at high energy and extrapolation
of these data down to the threshold, based on the numerical solutions
to the Roy equations of \cite{bfp}. 
It is obviously more efficient, as far as errors are concerned, 
to express directly the LEC's as   sum rules. Concerning the P-wave, 
the results found in the chiral expansion for the threshold parameters
$a_1^1$, $b_1^1$ and $c_1^1$ tend to support the idea that the P-wave
has a nearly pure Breit-Wigner shape down to the threshold. This indeed
implies the following relations:
\be
a_1^1={4\Gamma_V M^2_V\over(M^2_V-4)^{5\over2}},\quad
b_1^1={4a_1^1\over(M^2_V-4)},\quad c_1^1={4b_1^1\over(M^2_V-4)}\ ,
\en
where the mass and the width are expressed in units of the pion mass. One
can check that these relations are 
rather well satisfied for $a_1^1$ and $b_1^1$
and remain qualitatively correct even for $c_1^1$. 
An evaluation of $b_1^1$ on the basis of sum rules was performed 
only recently as an outcome of new, rapidly convergent 
sum rules involving combinations
of threshold parameters \cite{ananth} and the result is 
$b_1^1=(6\pm4)\times10^{-3}$. 

\medskip
\begin{center}
\begin{tabular}{|c|rrc|}\hline
\       &{1-loop}&{1-loop+2-loop}&Ref\cite{dumbrajs}\\ \hline
$10^3\,a_2^0$   & 1.52  &  1.72       &$ 1.7\pm0.3$ \\
$10^3\,a_2^2$   & 0.20  &  0.14       &$0.13\pm0.3$ \\
$10^3\,b_1^1$   & 4.09  &  5.46       &  $-$        \\ \hline
$10^4\,a_3^1$   & 0.30  &  0.58       &$0.6\pm0.2$  \\
$10^4\,b_2^0$   &$-4.79$  & $-3.41$       & $-$         \\
$10^4\,b_2^2$   &$-3.04$  & $-3.54$       & $-$         \\
$10^4\,c_1^1$   &$-1.82$  &  4.64       & $-$         \\ \hline
\end{tabular}
\vskip 0.5 truecm
{\bf Table 7:\ }{\sl\ Values of a few threshold parameters at one-loop
and at two-loop accuracy (in units of $M_{\pi^+}$) for $\alpha=2$
and $\beta=1.08$. The threshold parameters in the last four lines have no
tree-level contributions at order $O(p^4)$.}
\end{center}

It is instructive to compare our
results for the S and P partial-wave amplitudes 
derived in the two-loop approximation with a direct 
numerical solution of the Roy equations. This comparison 
allows to gauge the importance of $O(p^8)$ terms which are
present in the numerical solution and also allows to verify
whether the driving terms agree. Numerical results for the 
phase shifts have
been tabulated by Froggatt and Petersen \cite{fp} corresponding
to $a_0^0=0.30$ and $a_0^2=-0.018$. Using the chiral expansion
of the low-energy parameters (see appendix D of I) and
the sum rules, we obtain the following numbers for the
chiral parameters corresponding to
these scattering lengths: $\alpha=2.84$, $\beta=1.09$, $10^3\lambda_1=
-6.32,\ 10^3\lambda_2=9.77,\ 10^4\lambda_3=2.20$ and $10^4\lambda_4=
-1.46$.
Figure 1 shows
that the phase shifts computed from the two-loop chiral
expansion agree rather well with the results of \cite{fp} up
to an energy $\sqrt{s}\simeq 500$ MeV, thereby justifying
the assumption made in the derivation of the sum rules to
trust the chiral representation precisely in this range.

\bigskip
\leftline{\bf 3.5 Implications for the standard $\chi$PT two-loop amplitude
parameters}

\bigskip
While the present article was being completed, the computation of
the \pp scattering amplitude to two-loop accuracy has been achieved in the
framework of the standard \chpt\ in Ref.\cite{bij}. Comparing the 
expression of $A(s\vert t,u)$ obtained by the authors of 
\cite{bij} with the one we had derived in I, we conclude that they
coincide up to order $O(p^8)$ contributions, provided one identifies 
the six constants $b_i$ which appear in the amplitude of \cite{bij}
with our parameters $\alpha,\ \beta,\ \lambda_i$ as follows
\footnote{We are
indebted to G. Colangelo for pointing out a mistake in the first version of the
manuscript.}:
\bea
&&b_3=\lambda_1+{1\over2}\lambda_2-{\M^2\over\fpi^2}\left(6\lambda_3
        +{1\over12288\pi^2}-{103\over4608\pi^4}\right)
\nonumber\\
&&b_4={1\over2}\lambda_2+{{\M^2}\over{1536\pi^4\fpi^2}}
\nonumber\\
&&b_5=\lambda_3-{1\over4}\lambda_4-{19\over 1536\pi^4}
\nonumber\\
&&b_6=-{3\over4}\lambda_4-{7\over{4608\pi^4}}
\nonumber\\
&&b_1={\fpi^2\over3\M^2}(\alpha-1)-{4\over3}{\fpi^2\over\M^2}(\beta-1)+
4\lambda_1-{\M^2\over\fpi^2}\left(8\lambda_3 + {1\over{1152\pi^2}} -
{1\over{144\pi^4}}\right)
\nonumber\\
&&b_2={\fpi^2\over\M^2}(\beta-1)-4\lambda_1+{\M^2\over\fpi^2}
\left( 12\lambda_3\ +{19\over{18432\pi^2}}-{1\over{96\pi^4}}\right).
\ena
From the last two formulae, it is apparent that in the standard framework, 
both $\alpha$ and $\beta$ have to become equal to one in the chiral limit. (In
$G\chi$PT, $\alpha$ stays away from 1 and $b_1$ becomes large, signaling the
breakdown of the standard expansion.)

From our preceding determination of the four constants $\lambda_i$, we
can now deduce the values of the four constants $b_3,\ b_4,\ b_5$ and
$b_6$. In order to make the result as accurate as possible, the $O(p^6)$
values of $\alpha$ and $\beta$ in the standard framework would be 
required. Unfortunately, we see no way to extract these values from the
content of \cite{bij}. However, we expect that $\alpha$ and $\beta$ will
remain close to their leading order values $\alpha=\beta=1$ in the 
standard framework. The $O(p^4)$ values (in the two flavour standard $\chi$PT) 
$\alpha=1.06$, $\beta=1.10$
provide already some information about the expected corrections. Using
these values, we obtain (taking $F_\pi=93.2$ MeV here in order to conform
with \cite{bij}):
\bea
&&b_3= (-3.7\pm2.4)\,10^{-3}\qquad b_4=(4.8\pm0.3)\,10^{-3}\nonumber\\
&&b_5= ( 1.4\pm0.6)\,10^{-4}\ \ \qquad b_6=(1.0\pm0.1)\,10^{-4}
\ena
If one assumes that the $O(p^6)$ corrections to $\alpha$ and $\beta$
are of the order of the square of the $O(p^4)$ corrections, then the above 
numbers are not modified.

\newpage

\leftline{\large\bf 4. Summary and conclusions}

\bigskip
i) ~Among the six parameters $\alpha,\ \beta,\ \lambda_1,...,\lambda_4$
that define the two-loop \pp scattering amplitude, the parameter 
$\alpha$ is the most sensitive to the chiral structure of the QCD
vacuum. Furthermore, it is relatively weakly affected by higher orders
of \chpt, less affected than, for instance, the S-wave scattering lengths.
For these reasons, the parameter $\alpha$ represents a suitable 
quantitative characteristic of the strength of quark condensation,
and it should be determined experimentally in order to disentangle the
large and small condensate alternatives. To accomplish 
this, one needs new
high-precision low-energy \pp scattering data {\it and} a model-independent 
determination of the parameters $\lambda_1,...,\lambda_4$. Indeed, 
this would allow a measurement of $\alpha$ (and $\beta$) from a simultaneous
fit to several low-energy observables such as 
$\delta_0^0-\delta_1^1$ \cite{franzini} and $a_0^0-a_0^2$ \cite{dirac}. The
present paper contributes to this program by providing a rather accurate
determination of the parameters $\lambda_1,...,\lambda_4$ using existing
\pp scattering data at medium energies. 

ii) ~The method of determination follows from a rather systematic 
procedure: first, one establishes a dispersive representation for the 
low-energy amplitude in terms of absorptive parts and two unknown
subtraction {\it constants}. This representation is displayed in 
Eq.\rf{Adisp}, and it holds up to corrections of order $O[(p/E)^8]$, 
with $4\M^2<<E^2\lapprox 1$ GeV$^2$. Its derivation parallels the derivation 
of the Roy equations, using twice-subtracted fixed-t dispersion relations
and exploiting crossing symmetry. The second step consists in identifying, 
up to and including $O(p^6)$ accuracy, the dispersive representation 
\rf{Adisp} and the perturbative two-loop amplitude in a whole low-energy
domain of the Mandelstam plane. The comparison yields six, and only six,
sum rules for $\alpha,\ \beta,\ 
\lambda_1,\ \lambda_2,\ \lambda_3,\ \lambda_4$. The sum rules for $\alpha$
and $\beta$ are only barely convergent and they were ignored. One remains
with four rapidly convergent sum rules for $\lambda_1,...,\lambda_4$. Our
method minimizes the error in the determination of these parameters, 
since it makes full use of the information contained in crossing symmetry.

iii) ~In evaluating the sum rules \rf{sumrule}, one uses the existing 
\pp data in the energy range $0.5\ {\rm GeV}<\sqrt{s}<1.9$ GeV which 
were extracted from unpolarized high statistics $\pi N\to\pi\pi N$
high-energy production experiments. 
The outcome is rather sensitive to these data; in particular, their error bars
constitute the main source of uncertainty in the values of the 
$\lambda_i$'s (see table 6). Our machinery is ready to accept as input
any other set of medium energy \pp phases and inelasticities, provided
they are consistent with the Roy-type dispersion relations. (The standard
\pp phases extracted from the old Cern-Munich experiment \cite{grayer}
have been 
recently criticized \cite{svec}. However, no alternative phases have
been proposed.) The high-energy tails in the sum rules \rf{sumrule} are
estimated using the Regge-pole model and they are found to be negligible. 
Finally, the contribution of the low-energy range $2\M\le\sqrt{s}\le0.5$ 
GeV is evaluated using the perturbative two-loop amplitude itself. This
contribution introduces a weak dependence of the parameters $\lambda_1,...
\lambda_4$ on $\alpha$ and $\beta$ illustrated in table 4. This low-energy
part of the sum rules encompasses the bulk of infrared contributions
represented by the chiral logarithms in an explicit \chpt\ calculation. 

iv) ~Our final result can be read off from tables 3, 4 and 6. The variation
of the $\lambda_i$'s with the cutoff $E$ exhibits a plateau for $E\sim
700-900$ MeV. Choosing for the central values $E=800$ MeV as well as
the center of the variations with respect to $\alpha$ and to $\beta$, 
we obtain as the 
final result of this paper
\bea\lbl{final}
&&10^3\lambda_1=-6.1\pm2.2\ ,\qquad 10^3\lambda_2= 9.6\pm0.5\nonumber\\
&&10^4\lambda_3= 2.2\pm0.6\ ,\ \ \qquad 10^4\lambda_4=-1.45\pm0.12\ .
\ena
The central values may be further refined taking into account the 
small dependence on $\alpha$ and on $\beta$. In any case, the variation with
$\alpha$ and $\beta$ is smaller than the error bars displayed in 
\rf{final}. Giving to the parameters $\lambda_1$,...,$\lambda_4$ the
values \rf{final}, the two-loop perturbative amplitude becomes a faithful
{\it analytic} low-energy representation of the numerical solution 
of the Roy equations,  which were used to establish the values, usually
quoted as \lq\lq experimental", 
of \pp threshold parameters \cite{dumbrajs}. Indeed, for each pair of values of
$\alpha$ and $\beta$ ( $a_0^0$ and $a_0^2$ ) the perturbative amplitude
reproduces the S-wave slopes, and the P, D and F-wave threshold parameters as 
quoted in \cite{dumbrajs}.

v) ~The parameters $\lambda_1,...,\lambda_4$ are determined globally, 
without any reference to their expansion in powers of quark masses and
chiral logarithms. Consequently, the resulting values \rf{final} concern
both the standard and the generalized \chpt\ (modulo a small variation
with $\alpha$ and $\beta$ displayed in table 4). It would be interesting 
to see, whether the \lq\lq chiral anatomy" of these parameters resulting
from the recent standard \chpt\ two-loop calculation \cite{bij} can
be used to predict the values of $\lambda_1,...,\lambda_4$
(or, equivalently, $b_3,...,b_6$) in agreement with their present 
determination.
 
\section*{Acknowledgments}

This work was supported in part by the EEC Human Capital and Mobility
Progam, EEC-Contract No. CHRX-CT920026 (EURODA$\Phi$NE).  Useful
discussions and/or correspondance 
with B. Ananthanarayan, J. Bijnens, G. Colangelo,  G. Ecker, J. Gasser, J.
Kambor, D. Krupa, H. Leutwyler, R. Longacre, D. Morgan, M. R.
Pennington, J. Portol\'es, M. Svec and D. Toublan are greatly acknowledged.

\newpage
\begin{figure}[h]
\begin{center}
\leavevmode
\hbox{%
\epsfxsize=15.20 truecm\epsfbox{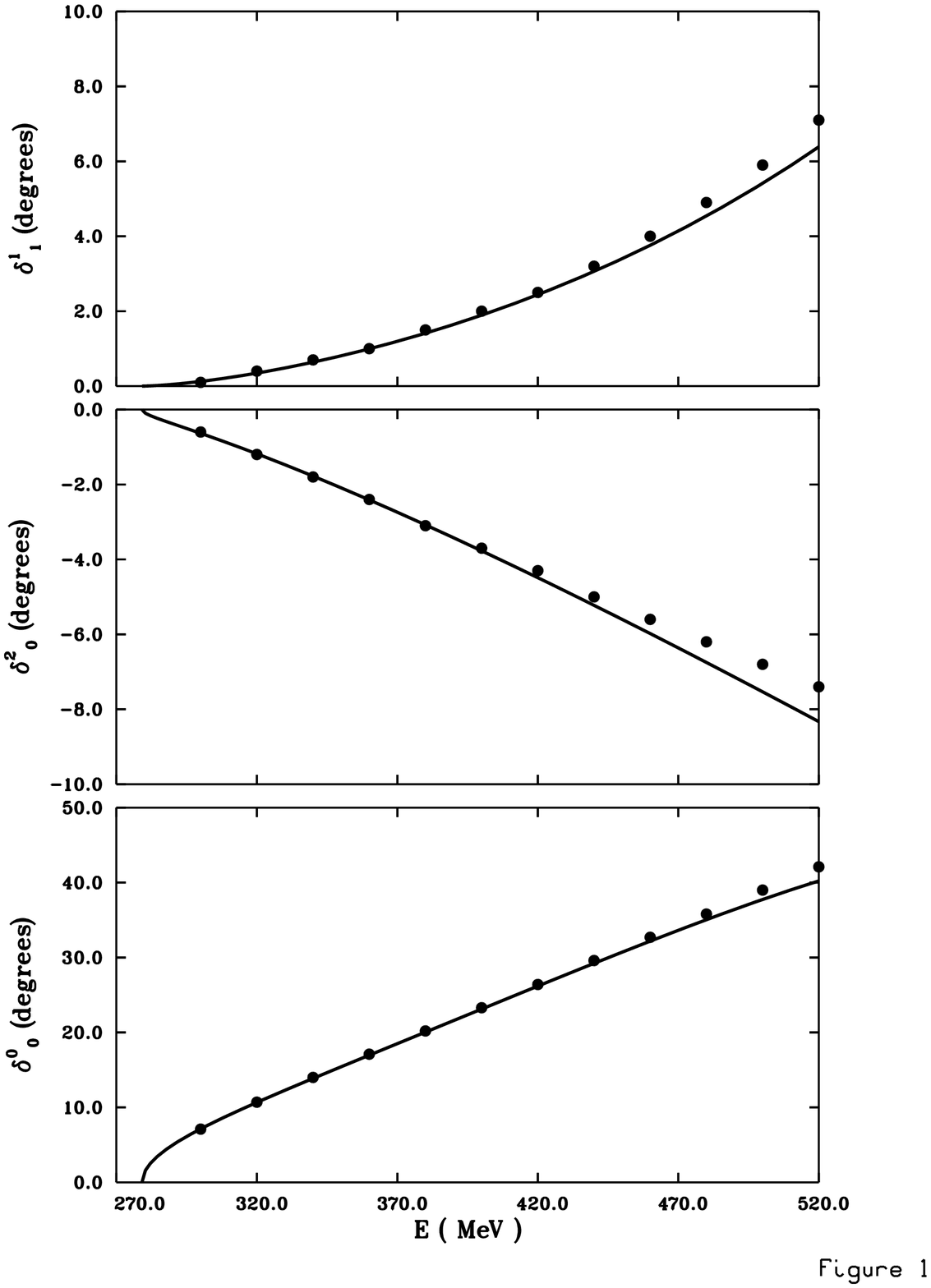} }
\vskip 0.5 truecm
{\bf Figure 1:}{Comparison of the S- and P- wave phase shifts
in the chiral expansion at two-loop with the results of a 
numerical solution
of the Roy equations quoted by Froggatt and Petersen \cite{fp} } 
\end{center}
\end{figure}
\end{document}